\documentclass[10pt]{article}
\usepackage{amsmath}
\usepackage{amssymb}
\usepackage{graphicx}

\usepackage{bm}

\begin{document}

\setcounter{figure}{0}
\setcounter{table}{0}
\setcounter{footnote}{0}
\setcounter{equation}{0}

\vspace*{0.5cm}

\noindent {\Large ENHANCED TERM OF ORDER $G^3$ IN THE LIGHT TRAVEL TIME:\\
DISCUSSION FOR SOME SOLAR SYSTEM EXPERIMENTS}
\vspace*{0.7cm}

\noindent\hspace*{1.5cm} P. TEYSSANDIER$^1$, B. LINET$^2$,   \\
\noindent\hspace*{1.5cm} $^1$ D\'ept SYRTE, CNRS/UMR 8630, UPMC,\\
\noindent\hspace*{1.5cm} Observatoire de Paris, 61 avenue de l'Observatoire, F-75014 Paris (France)\\
\noindent\hspace*{1.5cm} Pierre.Teyssandier@obspm.fr\\
\noindent\hspace*{1.5cm} $^2$ Laboratoire de Math\'ematiques et Physique Th\'eorique, CNRS/UMR 7350\\
\noindent\hspace*{1.5cm} Universit\'e Fran\c{c}ois Rabelais, F-37200 Tours (France)\\
\noindent\hspace*{1.5cm} Bernard.Linet@lmpt.univ-tours.fr\\

\vspace*{0.5cm}

\noindent {\large ABSTRACT.} It is generally believed that knowing the light travel time up to the post-post-Minkowskian level (terms in $G^2$) is sufficient for modelling the most accurate experiments designed to test general relativity in a foreseeable future. However, we have recently brought a rigorous justification of the existence of an enhanced term of order $G^3$ which becomes larger than some first-order contributions like the gravitomagnetic effect due to the rotation of the Sun or the solar quadrupole moment for light rays almost grazing the solar surface. We show that this enhanced term must be taken into account in solar system experiments aiming to reach an accuracy less than $10^{-7}$ in measuring the post-Newtonian parameter $\gamma$.

\vspace*{1cm}

\noindent {\large 1. INTRODUCTION}

\smallskip

In many experiments designed to test relativistic gravity, it is essential to calculate the light travel time $t_{\scriptscriptstyle B}-t_{\scriptscriptstyle A}$ between an emitter located at point ${\bm x}_{\scriptscriptstyle A}$ and a receiver located at point ${\bm x}_{\scriptscriptstyle B}$ as a function of ${\bm x}_{\scriptscriptstyle A}$ and ${\bm x}_{\scriptscriptstyle B}$ for a given time of reception $t_{\scriptscriptstyle B}$, namely the expression ${\cal T}_{r}({\bm x}_{\scriptscriptstyle A},t_{\scriptscriptstyle B}, {\bm x}_{\scriptscriptstyle B})$ such that
\begin{equation} \label{T}
t_{\scriptscriptstyle B}-t_{\scriptscriptstyle A} ={\cal T}_{r}(\bm x_{\scriptscriptstyle A},t_{\scriptscriptstyle B}, \bm x_{\scriptscriptstyle B}). 
\end{equation} 
The function ${\cal T}_{r}$ may be called the ``reception time transfer function". Knowing this function enables one to model the Doppler-tracking or the gravitational bending of light involved in the determinations of the post-Newtonian parameter $\gamma$ from solar system experiments (see, e.g., Le Poncin et al. 2004).

It is generally believed that projects like LATOR, ASTROD, SAGAS, ODYSSEY, GAME --- designed to measure the post-Newtonian parameter $\gamma$ at accuracies less than $10^{-7}$--- only require the determination of the light travel time up to the order $G^2$, with $G$ being the Newtonian gravitational constant (see, e.g., Minazzoli \& Chauvineau 2011 and refs. therein). However, this approach neglects the fact that some so-called ``enhanced" term of order $G^3$ in the time transfer function may become comparable to the regular term of order $G^2$, which can be estimated as const.$m^2/cr_c$, with $m$ being half the Schwarzschild radius of the central body and $r_c$ the 0th-order distance of closest approach of the light ray. The enhancement occurs in a close superior conjunction, i.e. in the case where the emitter and the receiver are almost on opposite sides of the central body---a configuration of crucial importance in experimental gravitation (Ashby \& Bertotti 2010). The third-order enhanced contribution can be recovered from the full expression of the time transfer function that we have recently obtained for a large class of static, spherically symmetric metrics generalizing the Schwarzschild solution (Linet \& Teyssandier 2013). We show that this contribution must be taken into account for modelling the above-mentioned experiments.

\vspace*{0.7cm}

\noindent {\large 2. TIME TRANSFER FUNCTION UP TO ORDER $G^3$}

\smallskip

The relativistic contributions to the light travel time due to the non-sphericity or to the motions of the Sun and the planets have been studied within the first order in $G$, and may be neglected beyond the linear regime (see, e.g., Klioner 1991, Kopeikin 1997, Linet \& Teyssandier 2002, Kopeikin \& Sch\"afer 1999, Kopeikin \& Mashoon 2002, Kopeikin et al. 2006, Zschocke \& Klioner 2011, Bertone et al. 2014, and refs. therein). So our investigation of the higher orders of approximation is confined to the static, spherically symmetric metrics describing the gravitational field of an isolated body of mass $M$. Spacetime is assumed to be covered by a single quasi-Cartesian coordinate system $x^{\mu}=(x^0, \bm x)$. For convenience, we put $x^0= ct$ and the spatial coordinates $\bm x$ are chosen so that the metric takes an isotropic form:
\begin{equation} \label{ds2}
ds^2={\cal A}(r)(dx^0)^2-{\cal B}(r)^{-1}d\bm x^2,
\end{equation}         
 where $r=\vert \bm x\vert$. This metric is considered as a generalization of the Schwarzschild metric. So it is assumed that ${\cal A}$ and ${\cal B}$ may be expanded in analytical series in $m/r$:
 \begin{eqnarray}
 && {\cal A}(r)=1 - \frac{2m}{r} + 2\beta \frac{m^2}{r^2} -\frac{3}{2}\beta_{3} \frac{m^3}{r^3}+\beta_4\frac{m^4}{r^4}+\sum_{n=5}^{\infty}\frac{(-1)^{n}n}{2^{n-2}} \beta_{n} \frac{m^n}{r^n}, \label{A}\\
&& {\cal B}(r)^{-1}=1 + 2 \gamma \frac{m}{r} +\frac{3}{2} \epsilon\frac{m^2}{r^2}+\frac{1}{2}\gamma_{3} \frac{m^3}{r^3}+\frac{1}{16}\gamma_4 \frac{m^4}{r^4} + \sum_{n=5}^{\infty}(\gamma_n-1)\frac{m^n}{r^n},\label{B}
\end{eqnarray}           
where $m=GM/c^2$ and the coefficients $\beta , \beta_3, \ldots , \beta_n, \ldots \gamma , \epsilon , \gamma_3 , \ldots \gamma_n \ldots$ are generalized post-Newtonian parameters chosen so that
\begin{equation} \label{ppN}
\beta = \beta_3 = \cdots = \beta_n = \cdots = 1, \quad \gamma = \epsilon = \gamma_3 = \cdots =\gamma_n=\cdots=1
\end{equation}
in general relativity.

Owing to the static character of the metric, the light travel time between $\bm x_{\scriptscriptstyle A}$ and $\bm x_{\scriptscriptstyle B}$ does not depend on the time of reception $t_{\scriptscriptstyle B}$ and equation (\ref{T}) reduces to $t_{\scriptscriptstyle B}-t_{\scriptscriptstyle A}={\cal T}(\bm x_{\scriptscriptstyle A}, \bm x_{\scriptscriptstyle B})$. In what follows, it is assumed that the time transfer function is expressible in a series in powers of $G$ having the form  
\begin{equation}  \label{expT}
{\cal T}(\bm x_{\scriptscriptstyle A}, \bm x_{\scriptscriptstyle B})=\frac{\vert \bm x_{\scriptscriptstyle B}-\bm x_{\scriptscriptstyle A}\vert}{c}+\sum_{n=1}^{\infty} {\cal T}^{(n)}(\bm x_{\scriptscriptstyle A}, \bm x_{\scriptscriptstyle B}),
\end{equation}
where ${\cal T}^{(n)}$ stands for the term of order $G^n$. Till lately, ${\cal T}$ was known only up to the second order in $G$; for ${\cal T}^{(1)}$, which is the well-known Shapiro time delay, see, e.g.,  Will 1993; for ${\cal T}^{(2)}$, see Le Poncin-Lafitte et al. 2004, Teyssandier \& Le Poncin-Lafitte 2008 and Klioner \& Zschocke 2010, which generalize the pioneering papers by John 1975, Richter \& Matzner 1983 and Brumberg 1987. Very recently, however, we have proposed a new procedure enabling one to determine ${\cal T}$ at any order of approximation (Linet \& Teyssandier 2013). Based on an iterative solution of an integro-differential equation derived from the null geodesic equations, this new procedure exclusively needs elementary integrations which may be performed with any symbolic computer program. We have obtained
\begin{eqnarray} 
& &{\cal T}^{(1)}( \bm x_{\scriptscriptstyle A}, \bm x_{\scriptscriptstyle B}) = \frac{(1+\gamma)m}{c}\ln\left(\frac{r_{\scriptscriptstyle A}+r_{\scriptscriptstyle B}+\vert \bm x_{\scriptscriptstyle B} -\bm x_{\scriptscriptstyle A} \vert}{r_{\scriptscriptstyle A}+r_{\scriptscriptstyle B}-\vert \bm x_{\scriptscriptstyle B} -\bm x_{\scriptscriptstyle A} \vert}\right), \label{T1} \\ 
& & \nonumber \\
& &{\cal T}^{(2)}( \bm x_{\scriptscriptstyle A}, \bm x_{\scriptscriptstyle B}) = \frac{m^2}{r_{\scriptscriptstyle A}r_{\scriptscriptstyle B}} \frac{\vert \bm x_{\scriptscriptstyle B} -\bm x_{\scriptscriptstyle A} \vert}{c}\bigg[ \kappa\frac{\arccos \bm n_{\scriptscriptstyle A}.\bm n_{\scriptscriptstyle B}}{\vert\bm n_{\scriptscriptstyle A}\times\bm n_{\scriptscriptstyle B}\vert}-
\frac{(1+\gamma)^2}{1+\bm n_{\scriptscriptstyle A}.\bm n_{\scriptscriptstyle B}}\bigg],\label{T2}\\ 
& & \nonumber \\
& &{\cal T}^{(3)}(\bm x_{\scriptscriptstyle A},\bm x_{\scriptscriptstyle B})=\frac{m^3}{r_{\scriptscriptstyle A}r_{\scriptscriptstyle B}}\left( \frac{1}{r_{\scriptscriptstyle A}}+\frac{1}{r_{\scriptscriptstyle B}}\right)\frac{\vert\bm x_{\scriptscriptstyle B}-\bm x_{\scriptscriptstyle A}\vert}{c(1+\bm n_{\scriptscriptstyle A}.\bm n_{\scriptscriptstyle B})} \bigg\lbrack \kappa_3 -(1+\gamma)\kappa\frac{\arccos \bm n_{\scriptscriptstyle A} .\bm n_{\scriptscriptstyle B}}{\vert\bm n_{\scriptscriptstyle A}\times\bm n_{\scriptscriptstyle B}\vert}+\frac{(1+\gamma )^3}{1+\bm n_{\scriptscriptstyle A}.\bm n_{\scriptscriptstyle B}}\bigg\rbrack, \label{T3} 
\end{eqnarray} 
where 
\begin{equation} \label{nAnB}
\bm n_{\scriptscriptstyle A}=\frac{\bm x_{\scriptscriptstyle A}}{r_{\scriptscriptstyle A}}, \qquad \bm n_{\scriptscriptstyle B}=\frac{\bm x_{\scriptscriptstyle B}}{r_{\scriptscriptstyle B}}
\end{equation}
and $\kappa$ and $\kappa_3$ are coefficients defined as
\begin{equation} \label{k23}
\kappa=2(1+\gamma)-\beta+\frac{3}{4}\varepsilon, \qquad \kappa_3=2\kappa-2\beta(1+\gamma)+\frac{1}{4}(3\beta_3 + \gamma_3).
\end{equation}
Of course, (\ref{T1}) and (\ref{T2}) coincide with the previously known results. On the other hand, (\ref{T3}) is new and enables us to determine the enhancement effects appearing in a superior conjunction up to order $G^3$.

\vspace*{0.7cm}


\noindent {\large 3. ENHANCED TERMS UP TO ORDER $G^3$}

\smallskip

In the case where $\bm x_{\scriptscriptstyle A}$ and $\bm x_{\scriptscriptstyle B}$ are in almost opposite directions (superior conjunction), an elementary geometrical reasoning shows that	
\begin{equation} \label{conj}
\frac{1}{1+\bm n_{\scriptscriptstyle A}.\bm n_{\scriptscriptstyle B}} \sim \frac{2 r_{\scriptscriptstyle A} r_{\scriptscriptstyle B}
}{(r_{\scriptscriptstyle A} +r_{\scriptscriptstyle B})^2}\frac{r_{\scriptscriptstyle A} r_{\scriptscriptstyle B}}{r_c^2},  
\end{equation}
where $r_c$ is the 0th-order distance of closest approach to the center of mass of the deflecting body, i.e. the Euclidean distance between the origin of the spatial coordinates and the line passing through $\bm x_{\scriptscriptstyle A}$ and $\bm x_{\scriptscriptstyle B}$:
\begin{equation} \label{rc}
r_c=\frac{r_{\scriptscriptstyle A}r_{\scriptscriptstyle B}\vert \bm n_{\scriptscriptstyle A}\times\bm n_{\scriptscriptstyle B} \vert}{\vert \bm x_{\scriptscriptstyle B} -\bm x_{\scriptscriptstyle A} \vert}.		
\end{equation} 
		
It follows from (\ref{conj}) that the first three perturbation terms in (\ref{expT}) are enhanced according to the asymptotic expressions (see Ashby \& Bertotti 2010 for a different method)
\begin{eqnarray}
&&{\cal T}^{(1)}_{enh}(\bm x_{\scriptscriptstyle A}, \bm x_{\scriptscriptstyle B})\sim \frac{(1+\gamma)m}{c}\ln\left(\frac{4 r_{\scriptscriptstyle A} r_{\scriptscriptstyle B}}{r^2_c}\right), \label{T1enh} \\ 
&&{\cal T}^{(2)}_{enh}(\bm x_{\scriptscriptstyle A}, \bm x_{\scriptscriptstyle B})\sim -2 \frac{(1+\gamma)^2 m^2}{c(r_{\scriptscriptstyle A}+r_{\scriptscriptstyle B})} \frac{r_{\scriptscriptstyle A} r_{\scriptscriptstyle B}}{r^2_c}, \label{T2enh} \\ 
&&{\cal T}^{(3)}_{enh}(\bm x_{\scriptscriptstyle A}, \bm x_{\scriptscriptstyle B})\sim 4\frac{(1+\gamma)^3 m^3}{c(r_{\scriptscriptstyle A}+r_{\scriptscriptstyle B})^2} \left(\frac{r_{\scriptscriptstyle A} r_{\scriptscriptstyle B}}{r^2_c}\right)^2. \label{T3enh}
\end{eqnarray}

The reliability of expansion (\ref{expT}) requires that inequalities 
\begin{equation} \nonumber
\left\vert{\cal T}^{(n)}( \bm x_{\scriptscriptstyle A}, \bm x_{\scriptscriptstyle B})\right\vert \ll \left\vert{\cal T}^{(n-1)}( \bm x_{\scriptscriptstyle A}, \bm x_{\scriptscriptstyle B})\right\vert  
\end{equation} 
are satisfied for any $n$, with ${\cal T}^{(0)}$ being conventionally defined as ${\cal T}^{(0)}(\bm x_{\scriptscriptstyle A}, \bm x_{\scriptscriptstyle B})=\vert \bm x_{\scriptscriptstyle B}-\bm x_{\scriptscriptstyle A}\vert/c $. It may be easily inferred from (\ref{T1enh})-(\ref{T3enh}) that these inequalities are fulfilled for $n=1, 2, 3$ as long as the distance of closest approach meets a condition as follows (see also Ashby \& Bertotti 2010):
\begin{equation} \label{vas}
\frac{2 m}{r_{\scriptscriptstyle A}+r_{\scriptscriptstyle B}} \frac{r_{\scriptscriptstyle A} r_{\scriptscriptstyle B}}{r^2_c} \ll 1.
\end{equation} 

Condition (\ref{vas}) is met in the currently designed solar system experiments. Indeed, assuming $r_{\scriptscriptstyle B}=1$ au and $r_{\scriptscriptstyle A}\geq r_{\scriptscriptstyle B}$, we have for light rays bypassing the Sun
\begin{equation} \label{vasS}
\frac{2 m_{\odot}}{r_{\scriptscriptstyle A}+r_{\scriptscriptstyle B}} \frac{r_{\scriptscriptstyle A} r_{\scriptscriptstyle B}}{r^2_c}\leq
9.12\times 10^{-4} \times \frac{R_{\odot}^2}{r_c^2},
\end{equation} 
where $R_{\odot}$ is the solar radius. 	

\vspace*{0.7cm}

\noindent {\large 4. APPLICATION TO SOLAR SYSTEM EXPERIMENTS}

\smallskip

1.	 Let us examine a SAGAS-like scenario (Wolf et al. 2009) aiming to determine the post-Newtonian parameter $\gamma$ at a level of accuracy reaching $10^{-8}$: $r_{\scriptscriptstyle A} \approx 50$ au, $r_{\scriptscriptstyle B} \approx 1$ au. Then Shapiro's formula (\ref{T1}) shows that ${\cal T}$ must be measured with an accuracy of $0.7$ ps (picosecond) in a superior conjunction configuration. Comparing this value with the contributions to the light travel time displayed in Table \ref{table1} shows that the enhanced term ${\cal T}_{enh}^{(3)}$ must be taken into account for rays almost grazing the Sun. 

Moreover, Table \ref{table1} shows that ${\cal T}_{enh}^{(3)}$ can be greater than the first-order gravitomagnetic effect $\vert{\cal T}^{(1)}_{S}\vert$ due to the spinning of the Sun and than the first-order contribution ${\cal T}^{(1)}_{J_2}$ due to the solar mass quadrupole.


\begin{table}[h]
\begin{center}
\begin{tabular}{l l l l l l}   \hline
\vspace{0.1 mm}$r_c/R_{\odot}$ & \vspace{0.1 mm}$\; \; | {\cal T}_{S}^{(1)} |$ & \vspace{0.1 mm}$\; \; \; \; {\cal T}_{J_2}^{(1)}$ & 
\vspace{0.1 mm}$\; \; \; \; \; {\cal T}_{enh}^{(2)}$ & \vspace{0.1 mm}$\; \; \; \; {\cal T}_{\kappa}^{(2)}$ & \vspace{0.1 mm}$\; \; \; \; {\cal T}_{enh}^{(3)}$ \\ \hline
$\; \; \; \; 1$ & $\; \; \; \; 10$ & $\; \; \; \; \; 2$ & $-17616$ & $\; \; 123$ & $\; \; \; \; 31.5$ \\
$\; \; \; \; 2$ & $\; \; \; \; \; \; 5$ & $\; \; \; \; \; 0.5$ & $\; \; -4404$ & $\; \; \; \; 61.5$ & $\; \; \; \; \; \; 2$ \\ 
$\; \; \; \; 5$ & $\; \; \; \; \; \; 2$ & $\; \; \; \; \; 0.08$ & $\; \; \; \; -704.6$ & $\; \; \; \; 24.6$ & $\; \; \; \; \; \; 0.05$ \\ \hline
\end{tabular}
\caption{Numerical values in ps of the light travel time \index{light travel time} in the solar system for various values of $r_c/R_{\odot}$. We put $r_{\scriptscriptstyle A}=50$ au, $r_{\scriptscriptstyle B}=1$ au, $\gamma=1$ and $\kappa =15/4$. For the numerical estimates of $\vert{\cal T}^{(1)}_{S}\vert$ and ${\cal T}^{(1)}_{J_2}$, the light ray is assumed to propagate in the equatorial plane of the Sun. For the solar quadrupole moment, we put $J_{2\odot}=2\times10^{-7}$ and for the internal angular momentum of the Sun, we take $S_{\odot}=2\times10^{41}$ kg m$^2$ s$^{-1}$ (see Komm et al. 2003). ${\cal T}_{\kappa}^{(2)}$ denotes the contribution due to $\kappa$ in the right hand side of (\ref{T2}).\label{table1}}
\end{center}
\end{table}

2. Consider now the deflection of light in a LATOR-like experiment, designed to reach an accuracy of a few $10^{-9}$ on $\gamma$ (Turyshev et al. 2009). The propagation direction of light is determined by the gradients of the function ${\cal T}$ (Le Poncin-Lafitte et al. 2004). For a ray passing near the Sun, the third-order enhanced term (\ref{T3enh}) yields a contribution $\Delta\chi^{(3)}_{enh}$ to the deflection beween the emitter and the receiver given by 	
\begin{equation} \label{ki3} 
\Delta\chi^{(3)}_{enh}\sim c\left\vert \bm \nabla_{\bm x_{\scriptscriptstyle A}}{\cal T}^{(3)}_{enh}(\bm x_{\scriptscriptstyle A}, \bm x_{\scriptscriptstyle B})+\bm \nabla_{\bm x_{\scriptscriptstyle B}}{\cal T}^{(3)}_{enh}(\bm x_{\scriptscriptstyle A}, \bm x_{\scriptscriptstyle B})\right\vert\sim \frac{16(1+\gamma)^3m^3}{r_c^3}\frac{r_{\scriptscriptstyle A} r_{\scriptscriptstyle B}}{(r_{\scriptscriptstyle A}+ r_{\scriptscriptstyle B})^2}\frac{r_{\scriptscriptstyle A} r_{\scriptscriptstyle B}}{r_c^2} .
\end{equation}
This estimate gives for $r_{\scriptscriptstyle A}\approx 1$ au, $ r_{\scriptscriptstyle B}\approx 1$ au and $r_c\approx R_{\odot}$ :
\begin{equation} \label{ki3g}
\Delta\chi^{(3)}_{enh}\approx 3 \, \mu \mbox{as} .
\end{equation}
Such a contribution cannot be neglected in the discussion since determining $\gamma$ at the level $5\times10^{-9}$ requires to measure the light deflection with an accuracy about $0.01$ $\mu$as.

3. The same conclusion is valid for an astrometric mission like GAME aiming to reach a $10^{-7}$ level, or better, in measuring $\gamma$ (Gai et al., 2012). The limit of (\ref{ki3}) when $r_{\scriptscriptstyle A} \rightarrow \infty$ yields a third-order contribution to the deflection of a light ray coming from infinity and observed at $\bm x_{\scriptscriptstyle B}$ given by 
\begin{equation} \label{ki3i}
\Delta\chi^{(3)}_{enh}\sim\frac{16(1+\gamma)^3m^3}{r_c^3}\left(\frac{r_{\scriptscriptstyle B}}{r_c}\right)^2.
\end{equation}
Taking $r_{\scriptscriptstyle B}\approx 1$ au and $r_c\approx R_{\odot}$, we find a contribution larger than the expected precision of $0.2$ $\mu$as :
\begin{equation} \label{ki3ig} 
\Delta\chi^{(3)}_{enh}\approx 12 \, \mu \mbox{as}.
\end{equation}
It may be noted that (\ref{ki3ig}) is in good agreement with the numerical estimate obtained in Hees et al. 2013.

\vspace*{0.7cm}

\noindent {\large 5. CONCLUSION}

Our explicit calculation of the time transfer function up to order $G^3$ for a large class of parametrized static, spherically symmetric metrics enables us to determine the enhanced contributions in the configurations of superior conjunction. It may be concluded that the third-order enhanced term given by (\ref{T3enh}) must be taken into account for modelling the future measurements of $\gamma$ from solar system experiments.

\smallskip

\vspace*{0.7cm}

\noindent {\large 6. REFERENCES}
%
%
%
%
%
{

\leftskip=5mm
\parindent=-5mm

\smallskip

Ashby, N., Bertotti, B., 2010, Class. Quantum Grav., 27, 145013 (27 pp).

Bertone, S., et al., 2014, Class. Quantum Grav., 31, 015021 (13 pp).

Brumberg, V. A., 1987 Kinematics Phys. Celest. Bodies, 3, pp. 6-12.

Gai, M., et al., 2012, Exp. Astron., 34, pp. 165-180.

Hees, A., Bertone, S., Le Poncin-Lafitte, C., 2013, Journ\'ees 2013 ``Syst\`emes de r\'ef\'erence spatio-temporels". 

John, R. W., 1975, Exp. Tech. Phys., 23, pp. 127-140.

Klioner, S. A., 1991, Sov. Astron., 35, pp. 523-530.

Klioner, S. A., Zschocke, S., 2010, Class. Quantum Grav., 27, 075015 (25 pp).

Komm, R., Howe, R., Durney, B. R., Hill, F., 2003, Astrophys. J., 586, pp. 650-662.

Kopeikin, S. M., 1997, J. Math. Phys., 38, pp. 2587-2601.

Kopeikin, S. M., Sch\"afer, G., 1999, Phys. Rev. D, 60, 124002 (44 pp).

Kopeikin, S. M., Mashhoon, B., 2002, Phys. Rev. D, 65, 064025 (20 pp).

Kopeikin, S. M., Korobkov, P., Polnarev, A., 2006, Class.Quantum Grav., 23, pp. 4299-4322.

Le Poncin-Lafitte, C., Linet, B., Teyssandier, P., 2004 Class. Quantum Grav., 21, pp. 4463-4483.

Le Poncin-Lafitte, C., Teyssandier, P., 2008, Phys. Rev. D, 77, 044029 (7 pp).

Linet, B., Teyssandier, P., 2002, Phys. Rev. D, 66, 024045 (14 pp).

Linet, B., Teyssandier, P., 2013, Class. Quantum Grav., 30, 175008 (23 pp).

Minazzoli, O., Chauvineau, B., 2011, Class. Quantum Grav., 28, 085010 (27 pp).

Richter, G. W., Matzner, R. A., 1983 Phys. Rev., D 28, pp. 3007-3012.

Teyssandier, P., Le Poncin-Lafitte, C., 2008, Class. Quantum Grav. 25, 145020 (12 pp).	

Turyshev, S. G., et al., 2009, Exp. Astron., 27, pp. 27-60.

Will, C. M., 1993, Theory and Experiment in Gravitational Physics (Cambridge University Press)

Wolf, P., et al., 2009, Exp. Astron., 23, pp. 651-687.

Zschocke, S., Klioner, S. A., 2011, Class. Quantum Grav., 28, 015009 (11 pp).

}

\end{document}